\documentclass[prd,preprintnumbers,12pt]{revtex4}
\pdfoutput=1
\usepackage{epsfig}
\usepackage{graphicx}
\usepackage{bm}

\newcommand{\be}{\begin{equation}}
\newcommand{\dd}{\displaystyle}
\newcommand{\ee}{\end{equation}}
\newcommand{\bea}{\begin{eqnarray}}
\newcommand{\eea}{\end{eqnarray}}

\newcommand{\nn}{\nonumber}
\newcommand{\de}{\partial}
\def\nn{\nonumber}
\def\de{\partial}

 \def\slash#1{\setbox0=\hbox{$#1$}#1\hskip-\wd0\dimen0=5pt\advance
       \dimen0 by-\ht0\advance\dimen0 by\dp0\lower0.5\dimen0\hbox
         to\wd0{\hss\sl/\/\hss}}
\def\be{\begin{equation}}
\def\dd{\displaystyle}
\def\ee{\end{equation}}

\def\bea{\begin{eqnarray}}
\def\eea{\end{eqnarray}}

\def\7{\tilde}
\def\8{\hat}

 \def\slash#1{\setbox0=\hbox{$#1$}#1\hskip-\wd0\dimen0=5pt\advance
       \dimen0 by-\ht0\advance\dimen0 by\dp0\lower0.5\dimen0\hbox
         to\wd0{\hss\sl/\/\hss}}

\def\tx{\tilde x}

\def\tZ{\tilde Z}

\def\tP{\tilde P}
\def\tB{\tilde B}
\def\tx{\tilde x}

\def\tZ{\tilde Z}
\def\tM{\tilde M}

\def\ttheta{\tilde \theta}

\usepackage{color}
\newcommand{\ddx}[1]{\frac{\partial}{\partial #1}}
\usepackage{hyperref}
\newcommand{\doi}[2]{{\hypersetup{urlcolor=darkred}\href{http://dx.doi.org/#2}{#1}\hypersetup{urlcolor=blue}}}
\newcommand{\eprintgrqcold}[1]{{\href{http://arxiv.org/abs/#1}{[\texttt{#1}]}}}
\newcommand{\eprintN}[1]{{\href{http://arxiv.org/abs/#1}{[\texttt{#1 [hep-th]}]}}}
\newcommand{\eprintgrqc}[1]{{\href{http://arxiv.org/abs/#1}{[\texttt{#1 [gr-qc]}]}}}

\begin{document}

{\hfill {\rm ICCUB-19-003}}
\title{ { Contractions of the Maxwell algebra}}
\author{Andrea Barducci}\email{barducci@fi.infn.it} \affiliation{Department of Physics, University of Florence and INFN Via G. Sansone 1, 50019 Sesto Fiorentino (FI), Italy}
\author{Roberto Casalbuoni}\email{casalbuoni@fi.infn.it}
\affiliation{Department of Physics, University of Florence and INFN Via G. Sansone 1, 50019 Sesto Fiorentino (FI), Italy}
\author{Joaquim Gomis}\email{joaquim.gomis@ub.edu}\affiliation{Departament de F\'isica Qu\`antica i Astrof\'isica\\ and
Institut de Ci\`encies del Cosmos (ICCUB), Universitat de Barcelona\\ Mart\'i i Franqu\`es , ES-08028 Barcelona, Spain
}


\begin{abstract}
 We construct  all the possible non-relativistic, non-trivial, Galilei and Carroll k-contractions
also known as k-1 p-brane contractions of the Maxwell algebra in $D+1$ space-time  dimensions.
$k$ has to do with the number of space-time dimensions one is contracting.
For non-trivial solutions, we { mean}
 the ones with a non-abelian algebra of
the momenta. We find in both cases, Galilei and Carroll, { eight} non trivial solutions. We also study the electromagnetic properties of the solutions, defined according to  the scaling performed on the 
 {generators} present in the Maxwell algebra. We find that besides the electric and magnetic contractions studied in  the literature for $k=1$, { that are related to the magnetic and electric limit of the free Maxwell
equations,} there also exist contractions where the two types of { fields} are scaled in the same way.

\end{abstract}
\maketitle

\section{Introduction}

{
Relativistic particles coupled to a constant electro-magnetic (EM) field enjoy symmetries that extend
the usual Poincar\'e algebra symmetries \cite{Bacry:1970ye}, in this case
the Lorentz generators
$M_{\mu\nu}$ where $ \mu,\nu=0,1\cdots D$,
that preserve the given EM field span a two dimensional abelian group. 
Associating  degrees of freedom to the constant EM field, one obtains a non-central extension of the Poincar\'e algebra generated by $M_{\mu\nu}$ and
$P_\mu$,
 by an anti-symmetric tensor
generator $Z_{\mu\nu}$.
The resulting algebra has
been called the Maxwell algebra in \cite{Schrader:1972zd}.
The generators of space-time translations become non-commuting, with commutators given  by}
\be
[P_\mu,P_\nu]=iZ_{\mu\nu},\label{eq:1}
\ee
with $Z_{\mu\nu}$ commuting with all $P_\mu$'s and among themselves. Of course, being a tensor the charges $Z_{\mu\nu}$ do not commute with the Lorentz generators:  {they transform covariantly under the Lorentz group.}
Therefore these charges are central for the translation group but not for the Poincar\'e group. 
 Physically the Maxwell algebra 
 was introduced
  as the symmetry  of the covariant solutions of the
  Klein-Gordon field in a  homogeneous classical electromagnetic field. 
  The Maxwell group has been studied further, for example, in 
  \cite{Beckers:1983gp,negro1,Soroka:2004fj}.
  The motion of a relativistic particle in a generic EM constant field, not fixed, was studied in \cite{Bonanos:2008ez} assuming the Maxwell group as the symmetry of the model. It was also noticed that there are infinite extensions
of the Maxwell algebra that are the symmetries of  a particle in a generic EM field.  The mathematical structure of these algebras has been elucidated in \cite{Gomis:2017cmt} as a Free Lie algebra generated by the space-time translations { 
generators $P_\mu$'s.}

The motion of a { non relativistic (NR)}  particle in a constant fixed EM  field was considered in 
\cite{Bacry:1970du}. The non-relativistic Maxwell algebras were studied in 
\cite{Beckers:1983gp,negro2,Bonanos:2008kr}. There are two types of NR Maxwell algebras that were
obtained as  magnetic  and electric limit of the EM field. { They are related to 
the two types of Galilean electromagnetism introduced in \cite{lebellac}. }
In an algebraic way   they correspond to  two type of inequivalent 
 { contractions of the Maxwell algebra. Contractions of Lie groups and Lie algebras were introduced in \cite{Inonu:1953sp,saletan}.
  }
 
{ Investigating the possible non-relativistic symmetry algebras is relevant for obtaining physical models with non-relativistic invariances. Beyond point particle dynamics, one can also
 consider realisations that lead to non-relativistic gravities 
 \cite{Cartan1,Cartan2,Trautman63,Havas:1964zza,DePietri:1994je,Andringa:2010it,Aviles:2018jzw,Hansen:2018ofj,Ozdemir:2019orp,Hansen:2019vqf,Bergshoeff:2019ctr,Aviles:2019xed}
or even symmetries of extended objects such as strings and branes 
\cite{Gomis:2000bd,Danielsson:2000gi,Gomis:2005pg,Brugues:2004an,Brugues:2006yd,
Batlle:2016iel}
}

In this paper we will study all the possible k-contractions { also known as $k-1$-brane contractions}
of the Maxwell algebra.
Here $k$ has to do with the number of space-time dimensions one is contracting, or scaling (see later for a more precise definition). We will be interested only in solutions leading to a non abelian algebra of the momenta. In fact, from a physical point of view, only in this case the charges $Z_{\mu\nu}$ lead to an interpretation in terms of an EM field \cite{Bonanos:2008ez}. Therefore, we will define  the case where the algebra of the momenta is abelian as a trivial one. In the trivial cases the algebraic structure is 
that of the contracted Poincar\'e algebra plus a set of charges with various  behaviour with respect to the contracted boosts.

{ The magnetic and electric NR limits of the Maxwell equations studied in \cite{lebellac}
are related to}
to two particular cases of  the $k=1$ contraction. These two different cases of  non relativistic limit give rise  to two different ways of  performing the contraction,  the magnetic and the electric contractions. When the magnetic field scales faster { to infinity} than the electric, we speak of a magnetic non-relativistic limit. In the opposite case, one speaks of  electric  non-relativistic limit. We will examine $k$ contractions of the Galilei and of the { Carroll \cite{Levy-Leblond}
case. 
 The two cases differ because in the Galilei case we scale k space variables, whereas for Carroll we scale the time variable plus k-1 space variables}.
For a generic  $k$-contraction, in both cases we find 8 non trivial  inequivalent contractions.  With respect to the  magnetic and electric cases we find 3 magnetic and 2 electric solutions, whereas in the remaining 3, the electric and the magnetic fields are scaled in the same way. 
In the particular case of $k=1$  for the Galilei case there are  only  3  non trivial solutions.  These 3 cases are one 
electric, one magnetic and one with equal scaling. 

{The contracted Maxwell  algebras we found can have interesting extensions. If we would like to obtain these algebras by contraction we will need to enlarge the Maxwell algebra. In the case
of three dimensions for k=1 these were studied in \cite{Aviles:2018jzw} by introducing U(1) factors. In this paper we will not consider the
general situation.}

In  the Carroll case all the 8 solutions are present for $k=1$. However,  for  $k=D$, the situation is similar to Galilei for $k=1$, that is, only 3 solutions lead to a non-abelian algebra of the momenta. On the other hand, all these solutions are now of an electric type.

This paper is organised as follows: in Section II we resume our method to obtain the contractions of the Poincar\'e group (see 
\cite{Barducci:2018wuj}). Furthermore we introduce the most general  contractions (or scaling) of the charges $Z_{\mu\nu}$
compatible with the rotational invariance.  The contractions of the charges depend on three exponents. The determination of these three parameters is going to fix the non trivial inequivalent solutions for the contractions of the Maxwell algebra with no central charges, both in the Galilei and in the Carroll cases. In Section III we study the $k$-contractions of the Galilei type 
leading to non trivial contractions. In Section IIIA we study these contracted algebras in coordinate space giving the explicit expressions for the generators of the contracted generators. An analogous study is done in Sections IV and IVA for the Carroll case. Conclusions  are in Section V. In the Appendix A we discuss the equivalence of two possible contractions for the Galilei case, but the same argument can be applied to Carroll. In Appendix B we prove that the definitions used for the contracted charges are the most general compatible with the rotational invariance.

\section{Description of the $k$-contractions}

In this Section we will study the $k$-contractions of Galilei and Carroll type of the Maxwell algebra.
{ In the case of the Poincar\'e algebra this was done in \cite{Barducci:2018wuj}.}
The Maxwell algebra in $D+1$ space-time dimensions has $(D+1)^2$ generators: $M_{\mu\nu}, P_\mu, Z_{\mu\nu}$, with commutation relations \cite{Schrader:1972zd}
\bea
 \left[M_{\mu\nu},M_{\rho\sigma}\right]&=&i(\eta_{\mu\rho}M_{\nu\sigma}+\eta_{\nu\sigma}M_{\mu\rho}
 -\eta_{\mu\sigma}M_{\nu\rho}
  -\eta_{\nu\rho}M_{\mu\sigma}),
\nonumber \\
  \left[M_{\mu\nu},P_{\rho}\right]&=&i(\eta_{\mu\rho}P_{\nu}-\eta_{\nu\rho}P_{\mu}),
\nn\\
  \left[P_{\mu},P_{\nu}\right]&=& iZ_{\mu\nu}, \nonumber \\
 \left [M_{\mu\nu}, Z_{\rho\sigma}\right] &=& i (\eta_{\mu\rho} Z_{\nu\sigma}+\eta_{\nu\sigma} Z_{\mu\rho} - \eta_{\nu\rho}Z_{\mu\sigma} - \eta_{\mu\sigma} Z_{\nu\rho} ),\nonumber \\
 \left [Z_{\mu\nu}, Z_{\rho\sigma}\right]&=& \left[Z_{\mu\nu},P_{\rho}\right]=0, 
  \label{eq:1.1}
\eea
with $\eta_{\mu\nu}=(-;+,\cdots,+)$ and $\mu,\nu=0,1,...,D$.

A property   that  will be useful  in the following is that the Maxwell algebra  
 is invariant under the following rescaling
\be
M_{\mu\nu}\to\alpha^0M_{\mu\nu},~~~P_\mu\to \alpha^1 P^\mu.~~~Z_{\mu\nu}\to \alpha^2Z_{\mu\nu},
\label{eq:3}
\ee
 the exponents of the scaling of the  generators correspond to the level of the generators in the Free Lie algebra description of Maxwell algebras \cite{Gomis:2017cmt}.
 The generator of this scaling is the generator $D$  of  the dilatations.

Since the momenta are not commuting,  the quadratic Casimir  $P^2$ is modified    \cite{Schrader:1972zd}, 
see also \cite{Soroka:2004fj} \cite{Gomis:2009dm}. Its expression is:
\be
C_2=P^2 +M_{\mu\nu} Z^{\mu\nu}.
\ee

In order to define the $k$ contractions of the Maxwell algebra we proceed as in 
\cite{Barducci:2018wuj}
 by partitioning the $D+1$ dimensional space-time in a $k$ dimensional Minkowskian part and in a $D+1-k$ dimensional Euclidean one 
 (for the case $k=1$ see also \cite{Beckers:1983gp,Aviles:2018jzw})   by introducing 
 the following set of labels for the space-time coordinates
\bea
&&\alpha,\beta =0,1,\cdots,k-1,~~~\eta_{\alpha\beta}=(-;+,\cdots,+),\nn\\
&&a,b =k,\cdots,D,~~~\eta_{ab}=(+,+,\cdots,+).
\eea

Let us recall how the $k$-contractions have been defined in \cite{Barducci:2018wuj,Barducci:2018thr} for the Poincar\'e case, (see
also \cite{Brugues:2004an,Gomis:2005pg,Brugues:2006yd}).
We have  to consider  the following two subgroups of $ISO(1,D)$: the Poincar\'e subgroup in $k$ dimensions, $ISO(1,k-1)$ and the euclidean group of roto-translations in $D+1-k$ dimensions, generated respectively by
\be
ISO(1,k-1):~~~M_{\alpha\beta},~~P_\alpha,~~~\alpha,\beta =0,1,\cdots,k-1,
\ee
\be
ISO(D+1-k):~~~M_{ab},~~P_a,~~~a,b =k,\cdots,D.
\ee
In these notations the generators of $ISO(1,D)$ are
\be
ISO(1,D):~~~M_{\alpha\beta},~~~M_{ab},~~P_\alpha,~~~P_a,~~~M_{\alpha b}\equiv B_{\alpha b}.
\ee
 Note that the boosts $ B_{\alpha b}$ connect the two subalgebras.

In \cite{Barducci:2018wuj} we have considered  two types of contractions, both at the level of the Poincar\'e algebra and at the level of 
 the invariant vector fields
. These contractions generalise the Carroll \cite{Levy-Leblond, Bergshoeff:2014jla, Duval:2014uoa, Cardona:2016ytk}  and the Galilei algebras 
\cite{LL} \cite{Gomis:2000bd, Danielsson:2000gi, Brugues:2004an, Gomis:2005pg, Brugues:2006yd} \cite{Batlle:2016iel,Gomis:2016zur,Batlle:2017cfa}. We note that we will be interested in the cases of Galilei and Carroll symmetries with no central charges.

At the  Lie algebra level the contractions are made on the momenta and on the boosts as follows
\bea
{\rm Galilei} :\tilde P_a ={\dd \frac 1\omega}P_a,~~~
 \tB_{\alpha a}= {\dd \frac 1\omega} B_{\alpha a},\label{eq:7}
\eea       
\bea
{\rm Carroll} :\tilde P_\alpha ={\dd \frac 1\omega}P_\alpha,~~~
\tilde B_{\alpha a}= {\dd \frac 1\omega} B_{\alpha a}.\label{eq:8}
\eea
and taking the limit $\omega\to\infty$.  The tilde generators  will be the ones associated to the "non-relativistic" algebras. 
The resulting algebra is
\be
{\rm Galilei}:[\tilde B_{\alpha a},\tB_{\beta c}]=0,~~~[\tilde B_{\alpha a}, \tilde P_\beta] =i\eta_{\alpha\beta}\tP_a,~~~[\tilde B_{\alpha a}, \tilde P_b]=0,\label{eq:1.11}
 \ee
 \be
 {\rm Carroll} :[\tilde B_{\alpha a},\tilde B_{\beta b}]=0,~~~[\tilde B_{\alpha a}, \tilde P_\beta] =0,~~~[\tilde B_{\alpha a}, \tilde P_b]=-i\eta_{ab}\tilde P_\alpha.\label{eq:1.9}
 \ee

Since the Poincar\'e algebra is invariant under a global rescaling of the momenta, the previous definition of the contractions is equivalent to:
\be
{\rm Galilei} : \tilde P_\alpha =\omega P_\alpha, ~~~\tilde B_{\alpha a}= {\dd \frac 1\omega} B_{\alpha a},\label{eq:13}
\ee
\be
{\rm Carroll} : \tilde P_a =\omega P_a, ~~~\tilde B_{\alpha a}= {\dd \frac 1\omega} B_{\alpha a}.\label{eq:14}
\ee
In the Maxwell case we have seen  that the  algebra is invariant under the rescaling given in eq. (\ref{eq:3}). It follows that also in this case the contractions defined in eqs.  (\ref{eq:13}) and  (\ref{eq:14}) are equivalent to the ones in eqs.  (\ref{eq:7}) and  (\ref{eq:8}). An explicit proof for the Galilei case is given in Appendix A. This proof can be simply extended to the Carroll case

We will complete  the   $k$-contraction of the Maxwell algebra through the following definition of the contracted charges $Z_{\mu\nu}$ that will be used both in the Galilei and the Carroll case:
\be
\tZ_{ab} =\omega^t Z_{ab},~~~\tZ_{a\alpha}=\omega^r Z_{a\alpha},~~~\tZ_{\alpha\beta}=\omega^s Z_{\alpha\beta}. 
\label{eq:15}
\ee
{ This definition} of the contracted charges is unique if we want to preserve the covariance in the Minkowski and in the Euclidean sectors of the space-time.

 Let us note  that the possible values of the three exponents $t,r,s$ will fix the possible contractions for the Maxwell algebra.

In the following two Sections we will consider  the Galilei and the Carroll cases
separately.

\section{Classification of the $k$-contractions for the Galilei case}

In this Section we will determine the values of the exponents $t,r,s$  of eq. (\ref{eq:15}) leading to a finite contracted algebra in the limit $\omega\to\infty$.
 The relevant commutators to be considered to this end are the following:
\be
[\tP_a,\tP_b]=\frac i{\omega^2}  Z_{ab}=\frac i{\omega^{t+2}}\tZ_{ab}\to t+2\ge 0,\label{eq:16}
\ee\be
[\tP_a,\tP_\alpha]=\frac i\omega  Z_{a\alpha}=\frac i{\omega^{r+1}}\tZ_{a\alpha}\to r+1\ge 0,\label{eq:17}
\ee
\be
[\tP_\alpha,\tP_\beta]=iZ_{\alpha\beta} =\frac i{\omega^s}\tZ_{\alpha\beta}, \to s\ge 0,\label{eq:18}
\ee
where we have used (\ref{eq:13}) and (\ref{eq:15}).
Let us note that whenever one of the exponents$,t,r,s$ is strictly greater than the values obtained in the previous equations, the commutators of the corresponding momenta vanish  in the limit $\omega\to\infty$, that is
\bea
&&t>-2 \to [\tP_a.\tP_b]=0,\nn\\
&&r>-1\to [\tP_a.\tP_\alpha]=0.\nn\\
&&s> 0\to  [\tP_\alpha.\tP_\beta]=0.\label{eq:19a}
\eea
Therefore,  the contractions corresponding to all the values of the exponents $t,r,s$ greater than the previous values are trivial,  in the sense specified in the Introduction.

\begin{figure}[h]
\begin{center}
   \includegraphics[width=6in]{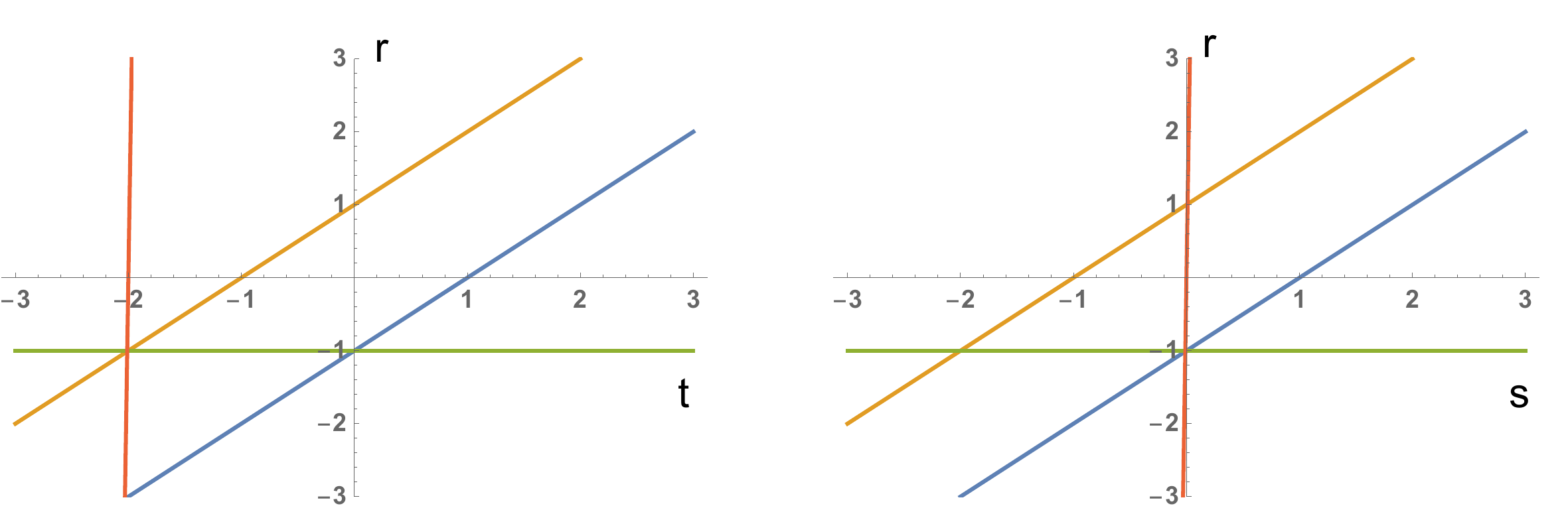} 
  \caption{\small The lines delimiting regions allowed for the contractions in the Galilei case. In the left-panel, $r$ vs. $t$. In the right one, $r$ vs. $s$.}
  \label{fig:1}
  \end{center}
  \end{figure}

Let us now consider the commutators of the boosts with the tensor charges. We have:

\be
[\tB_{\alpha a}, \tZ_{bc}]=\frac 1{\omega^{1-t}}[B_{\alpha a}, Z_{bc}]=\frac i{\omega^{1-t+r}}\left(\eta_{ac} \tZ_{\alpha b}-\eta_{ab} \tZ_{\alpha c}\right)\to 1-t+r\ge 0,\label{eq:19}
 \ee
\be
[\tB_{\alpha a}, \tZ_{b\beta}]=\frac 1{\omega^{1-r}} [B_{\alpha a}, Z_{b\beta}]=-\frac i{\omega^{1-r}}\left(\eta_{\alpha\beta}\omega^{-t}\tZ_{ab}+ \eta_{ab}\omega^{-s}\tZ_{\alpha\beta} \right)\to1 -r +t\ge 0, 1-r+s\ge 0,\label{eq:20}
\ee
\be
[\tB_{\alpha a}, \tZ_{\beta\gamma}]=\frac 1{\omega^{1-s}} [B_{\alpha a}, Z_{\beta\gamma}]=\frac i{\omega^{1-s}}\left(\eta_{\alpha\beta}\omega^{-r}\tZ_{a\gamma}- \eta_{\alpha\gamma}\omega^{-r}\tZ_{a\beta} \right)
\to 1+r-s\ge 0,\label{eq:21}
\ee
It follows:
\be
-1\le r-t\le +1,~~~-1\le r-s\le +1.\label{eq:23}
\ee
The allowed regions  for the parameters are shown in Fig. \ref{fig:1}, and are the ones to the right of the vertical lines, $t=-2$ and $s=0$, in  the upper part of the horizontal  line $r=-1$ and  in between the two diagonal lines:\\

 By inspection it is easy to find all the solutions leading to non vanishing commutators among all the momenta. We find 8 solutions that are listed in Table \ref{table:2}.
Note that these 8 solutions reduce to 3 non-trivial ones, in the case $k=1$, due to the vanishing of the tensor $\tZ_{\alpha\beta}$.
For all the other values of the triple $t,r,s$ the algebra of the momenta becomes abelian.

Of course, there are infinite trivial solutions with the tensor charges having different  commutation relations with the boosts  $\tB_{\alpha a}$ and with $\tM_{ab}$ and $\tM_{\alpha\beta}$. 
 
 Notice that  it is possible to obtain non-vanishing commutators of the tensor charges  with the boosts, whenever the parameters $t,r,s$ are at the borders of the regions inside the diagonal lines in Fig. \ref{fig:1}. 
  Without entering into a very detailed discussion, we list the four possibilities of having at least one commutator of the boosts with the charges different from zero:\\
\noindent
$r -t=-1$
\be
[\tB_{\alpha a}, \tZ_{bc}]=i\left(\eta_{ac} \tZ_{\alpha b} -\eta_{ab} \tZ_{\alpha c}\right)\equiv C^1_{\alpha a;bc}
\label{eq:24},
\ee
$r -t=+1,~ r-s\not = 1$ 
\be
[\tB_{\alpha a}, \tZ_{b\beta}]=-i\eta_{\alpha\beta}\tZ_{ab} \equiv C^2_{\alpha a;b\beta}            
\label{eq:25},
\ee
$r-s=+1, ~ r-t\not=1$
\be
[\tB_{\alpha a}, \tZ_{b\beta}]=-i\eta_{ab}\tZ_{\alpha\beta} \equiv C^3_{\alpha a;b\beta}
\label{eq:26},
\ee
$ r-s=-1$
\be
[\tB_{\alpha a}, \tZ_{\beta\gamma}]=i\left(\eta_{\alpha\beta}\tZ_{a\gamma}- \eta_{\alpha\gamma}\tZ_{a\beta} \right)
 \equiv C^4_{\alpha a;\beta\gamma}.
 \label{eq:27}
\ee
In Table \ref{table:2} we also list the relevant commutators for the 8 solutions of the exponents $t,r,s$.
\begin{table}[h]
\caption{The commutators for the inequivalent $k$ contractions of  Galilei type of the Maxwell algebra}
\begin{center}
\begin{tabular}{||c||c|c|c|c|c|c|c|c|c|c|c||}
\hline
$~n^0~$&~$t$~ &~$r$~ &~s$~$&$r-t$   & $r-s$   &$[\tP_{a},\tP_b]$ & $[\tP_\alpha,\tP_a]$ &$[\tP_\alpha,\tP_\beta]$ 
&$ [\tB_{\alpha a},\tZ_{bc}]$ & $ [\tB_{\alpha a},\tZ_{b\beta}]$  & $ [\tB_{\alpha a},\tZ_{\beta\gamma}]$        \\
\hline\hline
1&-2 & -1 & 0&+1   &-1     & $i\tZ_{ab}$ & $i\tZ_{\alpha a}$ & $i\tZ_{\alpha\beta}$ 
&0 &$C^2_{\alpha a;b\beta}$&$ C^4_{\alpha a;\beta\gamma}$\\
\hline
2&-1 & -1 & 0&0   &-1            & $0$ & $i\tZ_{\alpha a}$ & $i\tZ_{\alpha\beta}$ 
& 0& 0&$ C^4_{\alpha a;\beta\gamma}$\\
3&-1 & 0 & 0&+1   &  0   & $0$ & $0$ & $i\tZ_{\alpha\beta}$ 
& 0&$C^2_{\alpha a;b\beta}$ &0\\
\hline
4&0 & -1 & 0&-1   & -1    & $0$ & $i\tZ_{\alpha a}$ & $i\tZ_{\alpha\beta}$ 
&$C^1_{\alpha a;bc}$&0 &$ C^4_{\alpha a;\beta\gamma}$\\
5&0 & 0 & 0&0   &   0  & $0$ & $0$ & $i\tZ_{\alpha\beta}$ 
& 0& 0&0\\
6&0 & 1 & 0&  +1 &  +1   & $0$ & $0$ & $i\tZ_{\alpha\beta}$ 
& 0&$C^2_{\alpha a;b\beta}+C^3_{\alpha a;b\beta}$ &0\\
\hline
7&1 & 0 & 0   & -1    &0 & $0$& $0$ & $i\tZ_{\alpha\beta}$ 
&$C^1_{\alpha a;bc}$ &0 &0\\
8&1 & 1 & 0& 0  &  +1    & $0$& $0$ & $i\tZ_{\alpha\beta}$
& 0&$C^3_{\alpha a;b\beta}$ &0\\
\hline
\end{tabular}
\end{center}
\label{table:2}
\end{table}
where the coefficients $C^i$ are given in eqs. (\ref{eq:24})$-$(\ref{eq:27}).

In the case $k=1$ the only non trivial inequivalent contractions, in the sense that they lead to a non abelian algebra of the momenta, are those corresponding to the solutions 1), 2) and 4).

 We are now in the position to analyse in a more detailed way the so called magnetic and electric contractions that we discussed in the Introduction. In the case $k=1$, we have the following charges: $Z_{ab}$ and $Z_{a0}$ corresponding respectively to magnetic and electric fields. In the usual way of discussing the non relativistic limit, one scales the finite charges. In our case the finite charges are the tilde ones. Therefore we need to consider the limit $\omega\to\infty$ of the expressions
\be
Z_{ab} =\omega^{-t} \tZ_{ab},~~~~Z_{a0} =\omega^{-s} \tZ_{a0}.
\ee
We obtain the magnetic case when $-t>-r$ and the electric case in the opposite case. In other words, the magnetic and the electric case are discriminated by the values of $r-t$. By looking at Table \ref{table:2} we see that this quantity may assume three values, $\pm 1$ and $0$, magnetic case corresponding to $r-t =+1$ and the electric case to $r-t=-1$. However we see that another case is possible, namely the case where electric and magnetic field scale in the same way. This will be called the EM case.

By extension, also for $k\not =1$ we will define magnetic, electric and EM the cases corresponding to the three possible values of $r-t$. In the case $k=1$ we have only three non trivial contractions of all the three types. For a generic $k$, there are three magnetic solutions, two electric and three EM.

\subsection{$k$-contractions in configuration space}

We will now consider the $k$-contractions in the configuration  space spanned by  the coordinates of the coset space Maxwell/Lorentz,. The generic element of the coset space will be written as (see ref. \cite{Bonanos:2008ez})
\be\label{parametrization2}
g=e^{iP_\mu x^\mu}e^{\frac{i}{2}Z_{\mu\nu}\theta^{\mu\nu}}, 
\ee
with $\theta^{\mu\nu}=-\theta^{\nu\mu}$.
 Therefore, our configuration space will be parameterised by  
 $x^\mu$ and $\theta^{\mu\nu}$. The vector fields generating the Lorentz group transformations are given by:
\be
M_{\mu\nu}=i~(x_\mu \ddx {x^\nu}-x_\nu \ddx {x^\mu}+{\theta_\mu}^{\lambda}
\ddx {\theta^{\nu\lambda}}-{\theta_\nu}^{\lambda} \ddx {\theta^{\mu\lambda}}),\label{eq:34}\ee
whereas the  vector fields corresponding to $Z_{\mu\nu}$ and $P_\mu$ are (see ref. \cite{Bonanos:2008ez})
 \bea
Z_{\mu\nu}&=&-i~\ddx{\theta^{\mu\nu}},\nn\\
P_\mu&=&-i~(\ddx {x^\mu}-\frac{1}{2} x^\nu \ddx{\theta^{\mu\nu}}).\label{eq:35}.\eea
We note that the vector fields of eqs. (\ref{eq:34}) and  (\ref{eq:35}) are the so-called right-invariant vector fields generating the opposite Maxwell algebra, implying that the right hand side of the commutation relations has the opposite sign.

The contractions on the coordinates are obtained by the inverse scaling with respect to the corresponding generators, that is, in the Galilei case:
\be
\tx^a=\omega x^a,~~~\ttheta^{ab}=\omega^{-t}\theta^{ab},~~~
\ttheta^{\alpha a}=\omega^{-r}\theta^{\alpha a},~~~\ttheta^{\alpha\beta}=\omega^{-s}\theta^{\alpha\beta}\label{eq:32},
\ee
 Using eqs. (\ref{eq:34}), (\ref{eq:35}) and  (\ref{eq:32}) we find
\bea
\tZ_{\mu\nu}&=&-i~\ddx{\ttheta^{\mu\nu}},\\
\tP_a&=&-i~(\ddx {\tx^a}-\frac{1}{2(\omega^{2+t})} \tx^b \ddx{\ttheta^{ab}}-\frac 1{2(\omega^{1+r})}\tx^\beta\ddx{\ttheta^{a\beta}}),\\
\tP_\alpha&=&-i~(\ddx {\tx^\alpha}-\frac{1}{2(\omega^{1+r})} \tx^b \ddx{\ttheta^{\alpha b}}-\frac 1{2(\omega^{s})}\tx^\beta\ddx{\ttheta^{\alpha\beta}}),\\
\tB_{\alpha a}&=&i\left( \tx_\alpha\frac\de{\de \tx^a}-\frac{\ttheta_a^{.\,b}}{\omega^{1+r-t}}\frac{\de}{\de\ttheta^{\alpha b}}- \frac{\ttheta_a^{.\,\beta}}{\omega^{1-r+s}}\frac{\de}{\de\ttheta^{\alpha\beta}}+
\frac{\ttheta_\alpha^{.\,b}}{\omega^{1-r+t}}\frac{\de}{\de\ttheta^{ab}}+\frac{\ttheta_\alpha^{.\,\beta}}{\omega^{1-s+r}}\frac{\de}{\de\ttheta^{a\beta}} \right),\\
\tM_{\alpha\beta}&=&M_{\alpha\beta}(x\to\tx,\theta\to \ttheta),\\
\tM_{ab}&=&M_{ab}(x\to\tx,\theta\to\ttheta),\\
\eea

In the following Table \ref{table:3} we give the expressions of the generators of the contracted algebra for the 8 solutions we  previously found.

\begin{table}[h]
\caption{The generators of the contracted Maxwell algebras in  configuration  space}
\begin{center}
\begin{tabular}{||c||c|c|c|c|c|c|c|c||}
\hline
$~n^0~$&~$t$~ &~$r$~ &~s$~$&$r-t$   & $r-s$   &$\tP_{a}$ & $\tP_\alpha$ &$\tB_{a\alpha}$        \\
\hline\hline
1&-2 & -1 & 0&+1   &-1     & $-i~(\ddx {\tx^a}-\frac{1}{2} \tx^\mu \ddx{\ttheta^{a\mu}})$ & $-i~(\ddx {\tx^\alpha}-\frac{1}{2} \tx^\mu \ddx{\ttheta^{\alpha\mu}})$  & $i\left( \tx_\alpha\frac\de{\de \tx^a}+
\ttheta_\alpha^{.\,b}\frac{\de}{\de\ttheta^{ab}}+\ttheta_\alpha^{.\,\beta}\frac{\de}{\de\ttheta^{a\beta}} \right)$ 
\\

\hline
2&-1 & -1 & 0&0   &-1   &$-i~(\ddx {\tx^a}-\frac 1{2}\tx^\beta\ddx{\ttheta^{a\beta}})$& $-i~(\ddx {\tx^\alpha}-\frac{1}{2} \tx^\mu \ddx{\ttheta^{\alpha\mu}})$ & $i\left( \tx_\alpha\frac\de{\de \tx^a}+\ttheta_\alpha^{.\,\beta}\frac{\de}{\de\ttheta^{a\beta}} \right)$ 
\\
3&-1 & 0 & 0&+1   &  0   & $-i\ddx {\tx^a}$ & $-i(\ddx {\tx^\alpha}-\frac 1{2}\tx^\beta\ddx{\ttheta^{\alpha\beta}})$ & $i\left( \tx_\alpha\frac\de{\de \tx^a}+\ttheta_\alpha^{.\,b}\frac{\de}{\de\ttheta^{ab}} \right)$ 
\\
\hline
4&0 & -1 & 0&-1   & -1    & $-i(\ddx {\tx^a}-\frac 1{2}\tx^\beta\ddx{\ttheta^{a\beta}})$ & $-i~(\ddx {\tx^\alpha}-\frac{1}{2} \tx^\mu \ddx{\ttheta^{\alpha \mu}})$ & $i\left( \tx_\alpha\frac\de{\de \tx^a}-\ttheta_a^{.\,b}\frac{\de}{\de\ttheta^{\alpha b}}+\ttheta_\alpha^{.\,\beta}\frac{\de}{\de\ttheta^{a\beta}} \right)$ 
\\
5&0 & 0 & 0&0   &   0  & $-i\ddx {\tx^a}$ & $-i(\ddx {\tx^\alpha}-\frac 1{2}\tx^\beta\ddx{\ttheta^{\alpha\beta}})$ & $i\tx_\alpha\frac\de{\de \tx^a}$ \\
6&0 & 1 & 0&  +1 &  +1   & $-i\ddx {\tx^a}$ & $-i(\ddx {\tx^\alpha}-\frac 1{2}\tx^\beta\ddx{\ttheta^{\alpha\beta}})$ & $i\left( \tx_\alpha\frac\de{\de \tx^a}- \ttheta_a^{.\,\beta}\frac{\de}{\de\ttheta^{\alpha\beta}}+
\ttheta_\alpha^{.\,b}\frac{\de}{\de\ttheta^{ab}} \right)$ 
\\
\hline
7&1 & 0 & 0   & -1    &0 & $-i\ddx {\tx^a}$& $-i(\ddx {\tx^\alpha}-\frac 1{2}\tx^\beta\ddx{\ttheta^{\alpha\beta}})$ & $-i\left( -\tx_\alpha\frac\de{\de \tx^a}+\ttheta_a^{.\,b}\frac{\de}{\de\ttheta^{\alpha b}} \right)$ 
\\
8&1 & 1 & 0& 0  &  +1    & $-i\ddx {\tx^a}$& $-i(\ddx {\tx^\alpha}-\frac 1{2}\tx^\beta\ddx{\ttheta^{\alpha\beta}})$ & $i\left( \tx_\alpha\frac\de{\de \tx^a}- \ttheta_a^{.\,\beta}\frac{\de}{\de\ttheta^{\alpha\beta}} \right)$
\\
\hline
\end{tabular}
\end{center}
\label{table:3}
\end{table}

With this Table we conclude the classification of the $k$-contractions of the Maxwell algebra with no central charges for the Galilei case.
\newpage

\section{Classification of the $k$-contractions for the Carroll case}

We will now examine the Carroll case. The $k$-contraction is defined by eq. (\ref{eq:8})
\bea
\tilde P_\alpha ={\dd \frac 1\omega}P_\alpha,~~~
\tilde B_{\alpha a}= {\dd \frac 1\omega} B_{\alpha a}.\label{eq:2.7}
\eea
The other possibility of contractions as in eq. (\ref{eq:14}) can be shown to give
 the same results, following the same lines discussed in Appendix A for the Galilei case.

\begin{figure}[h]
\begin{center}
   \includegraphics[width=6in]{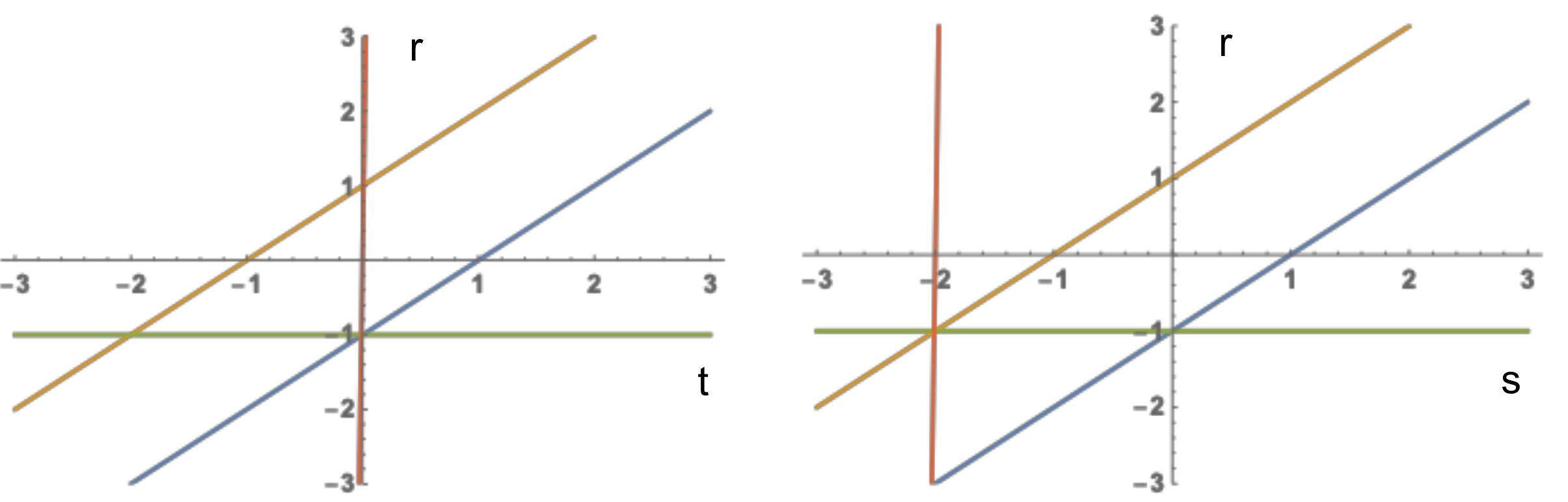} 
  \caption{\small The lines delimiting regions allowed for the contractions in the Carroll case. In the left-panel, $r$ vs. $t$. In the right one, $r$ vs. $s$.}
  \label{fig:2}
  \end{center}
  \end{figure}

The relevant commutation relations for the Carroll generators are
\be
[\tilde B_{\alpha a},\tilde B_{\beta b}]=0,~~~[\tilde B_{\alpha a}, \tilde P_\beta] =0,~~~[\tilde B_{\alpha a}, \tilde P_b]=-i\eta_{ab}\tilde P_\alpha.\label{eq:1.9}
 \ee
We  define the contracted tensor charges as in the Galilei case:
\be
\tZ_{ab} =\omega^t Z_{ab},~~~\tZ_{a\alpha}=\omega^r Z_{a\alpha},~~~\tZ_{\alpha\beta}=\omega^s Z_{\alpha\beta}. 
\ee
Again, in order to get a well defined contracted algebra we impose the following requirements:
\be
[\tP_a,\tP_b]= i  Z_{ab}=\frac i{\omega^{t}}\tZ_{ab}\to t\ge 0\label{eq:78},
\ee\be
[\tP_a,\tP_\alpha]=\frac i\omega  Z_{a\alpha}=\frac i{\omega^{r+1}}\tZ_{a\alpha}\to r+1\ge 0\label{eq:79},
\ee
\be
[\tP_\alpha,\tP_\beta]=\frac i{\omega^2}Z_{\alpha\beta} =\frac i{\omega^{s+2}}\tZ_{\alpha\beta}, \to s+2\ge 0\label{eq:80}.
\ee
As for the commutators of the Carroll boosts with the tensor charges we get the same results as in eqs. (\ref{eq:19}), (\ref{eq:20}) and (\ref{eq:21}):
\be
[\tB_{\alpha a}, \tZ_{bc}]=\frac 1{\omega^{1-t}}[B_{\alpha a}, Z_{bc}]=\frac i{\omega^{1-t+r}}\left(\eta_{ac} \tZ_{\alpha b}-\eta_{ab} \tZ_{\alpha c}\right)\to 1-t+r\ge 0,
 \ee
\be
[\tB_{\alpha a}, \tZ_{b\beta}]=\frac 1{\omega^{1-r}} [B_{\alpha a}, Z_{b\beta}]=-\frac i{\omega^{1-r}}\left(\eta_{\alpha\beta}\omega^{-t}\tZ_{ab}+ \eta_{ab}\omega^{-s}\tZ_{\alpha\beta} \right)\to1 -r +t\ge 0, 1-r+s\ge 0,\ee
\be
[\tB_{\alpha a}, \tZ_{\beta\gamma}]=\frac 1{\omega^{1-s}} [B_{\alpha a}, Z_{\beta\gamma}]=\frac i{\omega^{1-s}}\left(\eta_{\alpha\beta}\omega^{-r}\tZ_{a\gamma}- \eta_{\alpha\gamma}\omega^{-r}\tZ_{a\beta} \right)
\to 1+r-s\ge 0,
\ee
from which
\be
-1\le r-t\le +1,~~~-1\le r-s\le +1.
\ee
The allowed regions are shown in Fig. \ref{fig:2}, and are the ones to the right of the vertical lines, $t=0$ and $s=-2$, the one in  the upper part of the horizontal  line $r=-1$ and  in  between the two diagonal lines.\\
  
  From this we conclude that the non trivial solutions of the previous inequalities  for the exponents $t,r,s$ of this case, can be 
  obtained by looking at the solutions for the Galilei case with the exchange $s\leftrightarrow t$.  Therefore we can use the Table \ref{table:2} of the Galilei case, by performing the substitution $s$ with $t$. At the same time, it is clear 
  from eqs. (\ref{eq:78}) and  (\ref{eq:80})  that we need to perform  the exchange $
  \tZ_{ab} \leftrightarrow 
  \tZ_{\alpha\beta}$, and of the corresponding commutators among the momenta,  
  in  Table \ref{table:2}. Also, exchanging $s$ with $t$ the differences $r-t$ and $r-s$ are exchanged, implying a corresponding change in the commutators with the boosts, according to the eqs. (\ref{eq:24})$-$(\ref{eq:27}).  Taking everything into account, we obtain the Table   \ref{table:4}.


We note that contrarily to the Galilei case, where for $k=1$ only 3 non trivial solutions were surviving, in the Carroll case, for $k=1$ all the 8 solutions are non trivial. However. in the present case the non trivial solutions reduce to 3 for $k=D$, since then  $Z_{ab}=0$.

 As for the electric and magnetic behaviour, in the Carroll case there are 3 electric solutions, 2 magnetic and 3 EM. Therefore the number of the electric and magnetic solutions are exchanged with respect to the Galilei case.
In ref.  \cite{Duval:2014uoa}  only the electric and magnetic case were considered for $k=1$

  \begin{table}
\caption{The commutators for the inequivalent $k$ contractions of  Carroll type of the Maxwell algebra}
\begin{center}
\begin{tabular}{||c||c|c|c|c|c|c|c|c|c|c|c||}
\hline
$~n^0~$&~$s$~ &~$r$~ &~t$~$&$r-s$   & $r-t$   &$[\tP_{a},\tP_b]$ & $[\tP_\alpha,\tP_a]$ &$[\tP_\alpha,\tP_\beta]$ 
&$ [\tB_{\alpha a},\tZ_{bc}]$ & $ [\tB_{\alpha a},\tZ_{b\beta}]$  & $ [\tB_{\alpha a},\tZ_{\beta\gamma}]$        \\
\hline\hline
1&-2 & -1 & 0&+1   &-1     & $i\tZ_{ab}$ & $i\tZ_{\alpha a}$ & $i\tZ_{\alpha\beta}$ 
&$C^1_{\alpha a;bc}$ &$C^3_{\alpha a;b\beta}$&$ 0$\\
\hline
2&-1 & -1 & 0&0   &-1            & $i\tZ_{ab}$ & $i\tZ_{\alpha a}$ & $0$ 
& $C^1_{\alpha a;bc}$& 0&$0$\\
3&-1 & 0 & 0&+1   &  0   & $i\tZ_{ab}$ & $0$ & $0$ 
& 0&$C^3_{\alpha a;b\beta}$ &0\\
\hline
4&0 & -1 & 0&-1   & -1    & $i\tZ_{ab}$ & $i\tZ_{\alpha a}$ & $0$ 
&$C^1_{\alpha a;bc}$&0 &$ C^4_{\alpha a;\beta\gamma}$\\
5&0 & 0 & 0&0   &   0  & $i\tZ_{ab}$ & $0$ & $0$ 
& 0& 0&0\\
6&0 & 1 & 0&  +1 &  +1   & $i\tZ_{ab}$ & $0$ & $0$ 
& 0&$C^2_{\alpha a;b\beta}+C^3_{\alpha a;b\beta}$ &0\\
\hline
7&1 & 0 & 0   & -1    &0 & $i\tZ_{ab}$& $0$ & $0$ 
&$0$ &0 &$ C^4_{\alpha a;\beta\gamma}$\\
8&1 & 1 & 0& 0  &  +1    & $i\tZ_{ab}$& $0$ & $0$
& 0&$C^2_{\alpha a;b\beta}$ &0\\
\hline
\end{tabular}
\end{center}
\label{table:4}
\end{table}

\subsection{Configuration space}

The contracted variables in configuration space are:

\be
\tx^\alpha=\omega x^\alpha,~~~\ttheta^{ab}=\omega^{-t}\theta^{ab},~~~
\ttheta^{\alpha a}=\omega^{-r}\theta^{\alpha a},~~~\ttheta^{\alpha\beta}=\omega^{-s}\theta^{\alpha\beta}\label{eq:2.32},
\ee
and the expressions for the generators are:

\bea
\tZ_{\mu\nu}&=&-i~\ddx{\ttheta^{\mu\nu}},\\
\tP_a&=&-i~(\ddx {\tx^a}-\frac{1}{2(\omega^{t})} \tx^b \ddx{\ttheta^{ab}}-\frac 1{2(\omega^{1+r})}\tx^\beta\ddx{\ttheta^{a\beta}}),\\
\tP_\alpha&=&-i~(\ddx {\tx^\alpha}-\frac{1}{2(\omega^{1+r})} \tx^b \ddx{\ttheta^{\alpha b}}-\frac 1{2(\omega^{2+s})}\tx^\beta\ddx{\ttheta^{\alpha\beta}}),\\
\tB_{\alpha a}&=&i\left(- \tx_a\frac\de{\de \tx^\alpha}-\frac{\ttheta_a^{.\,b}}{\omega^{1+r-t}}\frac{\de}{\de\ttheta^{\alpha b}}- \frac{\ttheta_a^{.\,\beta}}{\omega^{1-r+s}}\frac{\de}{\de\ttheta^{\alpha\beta}}+
\frac{\ttheta_\alpha^{.\,b}}{\omega^{1-r+t}}\frac{\de}{\de\ttheta^{ab}}+\frac{\ttheta_\alpha^{.\,\beta}}{\omega^{1-s+r}}\frac{\de}{\de\ttheta^{a\beta}} \right),\\
\tM_{\alpha\beta}&=&M_{\alpha\beta}(x\to\tx,\theta\to \ttheta),\\
\tM_{ab}&=&M_{ab}(x\to\tx,\theta\to\ttheta),\\
\eea
  
  In   Table \ref{table:5} we give the expressions for  the generators in configuration space for the various solutions.   This concludes the classification of the $k$-contractions of the Maxwell algebra with no central charges in the Carroll case
  \begin{table}[h]
\caption{The generators of the contracted Maxwell algebras in configuration  space for the Carroll case}
\begin{center}
\begin{tabular}{||c||c|c|c|c|c|c|c|c||}
\hline
$~n^0~$&~$s$~ &~$r$~ &~t$~$&$r-s$   & $r-t$   &$\tP_{a}$ & $\tP_\alpha$ &$\tB_{a\alpha}$        \\
\hline\hline
1&-2 & -1 & 0&+1   &-1     & $-i~(\ddx {\tx^a}-\frac{1}{2} \tx^\mu \ddx{\ttheta^{a\mu}})$ & $-i~(\ddx {\tx^\alpha}-\frac{1}{2} \tx^\mu \ddx{\ttheta^{\alpha\mu}})$  & $ - i( \tx_a\frac\de{\de \tx^\alpha}+
{\ttheta_a^{.\,b}}\frac{\de}{\de\ttheta^{\alpha b}}+
{\ttheta_a^{.\,\beta}}\frac{\de}{\de\ttheta^{\alpha\beta}})$ 
\\
\hline
2&-1 & -1 & 0&0   &-1   &$-i~(\ddx {\tx^a}-\frac 1{2}\tx^\mu\ddx{\ttheta^{a\mu}})$& $ -i(\ddx {\tx^\alpha}-\frac{1}{2} \tx^b \ddx{\ttheta^{\alpha b}})$ & $-i\left( \tx_a\frac\de{\de \tx^\alpha}+
{\ttheta_a^{.\,b}}\frac{\de}{\de\ttheta^{\alpha b}} \right) $ 
\\
3&-1 & 0 & 0&+1   &  0   & $ -i~\left(\ddx {\tx^a}-\frac{1}{2} \tx^b \ddx{\ttheta^{ab}}\right)$ & $-i\ddx {\tx^\alpha}$ & $ i\left(- \tx_a\frac\de{\de \tx^\alpha}- 
{\ttheta_a^{.\,\beta}}\frac{\de}{\de\ttheta^{\alpha\beta}} \right)$ 
\\
\hline
4&0 & -1 & 0&-1   & -1    & $-i(\ddx {\tx^a}-\frac 1{2}\tx^\mu\ddx{\ttheta^{a\mu}})$ & $ -i~\left(\ddx {\tx^\alpha}-\frac{1}{2} \tx^b \ddx{\ttheta^{\alpha b}}\right)$ & $- i\left( \tx_a\frac\de{\de \tx^\alpha}+
{\ttheta_a^{.\,b}}\frac{\de}{\de\ttheta^{\alpha b}}-{\ttheta_\alpha^{.\,\beta}}\frac{\de}{\de\ttheta^{a\beta}} \right)$ 
\\
5&0 & 0 & 0&0   &   0  & $-i~\left(\ddx {\tx^a}-\frac{1}{2} \tx^b \ddx{\ttheta^{ab}}\right)$ & $ -i~\ddx {\tx^\alpha} $ & $ -i \tx_a\frac\de{\de \tx^\alpha}$ \\
6&0 & 1 & 0&  +1 &  +1   & $ -i~\left(\ddx {\tx^a}-\frac{1}{2} \tx^b \ddx{\ttheta^{ab}}\right)$ & $ -i\ddx {\tx^\alpha}$ & $i\left(- \tx_a\frac\de{\de \tx^\alpha}- 
{\ttheta_a^{.\,\beta}}\frac{\de}{\de\ttheta^{\alpha\beta}}+
{\ttheta_\alpha^{.\,b}}\frac{\de}{\de\ttheta^{ab}} \right) $ 
\\
\hline
7&1 & 0 & 0   & -1    &0 & $ -i~\left(\ddx {\tx^a}-\frac{1}{2} \tx^b \ddx{\ttheta^{ab}}\right)$& $-i\ddx {\tx^\alpha}$ & $-i\left( \tx_a\frac\de{\de \tx^\alpha}-
{\ttheta_\alpha^{.\,\beta}}\frac{\de}{\de\ttheta^{a\beta}} \right)$ 
\\
8&1 & 1 & 0& 0  &  +1    & $ -i~\left(\ddx {\tx^a}-\frac{1}{2} \tx^b \ddx{\ttheta^{ab}}\right)$& $-i\ddx {\tx^\alpha}$ & $i\left(- \tx_a\frac\de{\de \tx^\alpha}+
{\ttheta_\alpha^{.\,b}}\frac{\de}{\de\ttheta^{ab}} \right)
$
\\
\hline
\end{tabular}
\end{center}
\label{table:5}
\end{table}
  \newpage

   \section{Conclusions and outlook}
    In this paper we have studied the non trivial $k$-contractions of  the  relativistic Maxwell algebra.
     The  peculiarity of this algebra is to give rise to non commuting momenta,  which in physical terms correspond to the presence of  a constant EM field expressed by the right hand side of the momenta commutators. Therefore, we have defined as trivial all the contractions leading to  
abelian momenta. In both types of contractions, Galilei and Carroll, we have found 8 non trivial $k$-contractions. In the Galilei type of contractions for $k=1$ there are only 3 non trivial solutions due the fact that the charges $Z_{\alpha\beta}$ are vanishing. In the Carroll case this does not happen for $k=1$, but rather for $k=D$, in which case the charges $Z_{ab}$ are vanishing.  
   
 We have also studied the solutions from the point of  view of the electric and magnetic properties. Recalling that for $k=1$ the charges $Z_{ab}$ are associated with
  a magnetic field, whereas the charges $Z_{a0}$ with an electric one, we have followed the literature, defining these properties according to the difference $r-t$, where $r$ and $t$ are the exponents of the scaling of  $Z_{a\alpha}$ and $Z_{ab}$ respectively. More precisely  we call magnetic the solutions with a positive value of $r-t$. It turns out that the non trivial contractions have $r-t=\pm 1,0$, showing that besides the magnetic and electric contractions there is another type with the fields scaled in the same way. We have  called these solutions  the EM solutions.
 
  \begin{table}[h]
\caption{The non trivial $k$-contractions for Galilei and Carroll according to their magnetic and electric properties. In parenthesis the  non trivial solutions for $k=1$ in the Galilei case and for $k=D$ for Carroll.}
\begin{center}
\begin{tabular}{||c||c|c|c||}
\hline\hline
&~ Magnetic: $r-t=+1$~ & ~Electric:  $r-t=-1$~ &~ EM: $r-t=0$~\\
\hline\hline
Galilei &  (1), 3, 6 & (4), 7 & (2), 5, 8\\
\hline
Carroll & 6, 8 & (1, 2, 4) & 3, 5, 7\\
\hline\hline\end{tabular}
\end{center}
\label{table:6}
\end{table}

 In Table \ref{table:6} we summarise the solutions we found with respect to their electric and magnetic properties. In the Galilei case we have found 3 magnetic solutions, 2 electric and 3 EM. The situation  is somewhat inverted for Carroll. In fact, in this case we find 2 magnetic, 3 electric and 3 EM solutions. The solutions enclosed in parenthesis in the Table are the non trivial ones for $k=1$  for Galilei and $h=D$ for Carroll.  Whereas in the first case the three solutions  all have different  electromagnetic properties, being one magnetic, one electric and one EM. On the contrary, for Carroll all the 5 non trivial solutions for $k=D$ are of the electric type.
 
 For the future it would be interesting to  find the 
  extensions of the k-contracted algebras
 we have found in this  paper. { One should generalize the results of  the case k=1 in three dimensions 
\cite{Aviles:2018jzw}, where  the relativistic  Maxwell algebra was enlarged with U(1) factors}
 
 It will also be  interesting to perform the k-contractions of the Maxwell algebras of reference \cite{Bonanos:2008ez,Gomis:2017cmt}
 and compute their extensions.
 
{ From a physical point of view it will be interesting to construct NR gravities associated with the algebras we have found and look for NR dynamical objects that can be coupled to them. These objects could be useful to study strongly coupled systems in condensed matter using NR holography \cite{Sachdev} \cite{Liu}. Also the contracted algebras obtained from the Maxwell algebra, could be useful for building NR models of matter interacting with constant EM fields.}

 

   \section*{Acknowledgments}
   
   We acknowledge discussions with  Luis Avil\'es, Carles Batlle,   Diego Hidalgo, Axel Kleinschmidt, Jakob Palmkvist, Jakob Salzer  and Jorge Zanelli. 
JG has been supported in part by MINECO FPA2016-76005-C2-1-P and Consolider CPAN, and by the Spanish government (MINECO/FEDER) under project MDM-2014-0369 of ICCUB (Unidad de Excelencia Mar\'{i}a de Maeztu).    
 \section{Appendix A - Equivalent description of the Galilei type $k$-contractions}

Since the Poincar\'e algebra is invariant under a global rescaling of the momenta, the previous definition of the contractions is equivalent to:
\be
{\rm Galilei} : \tilde P_\alpha =\omega P_\alpha, ~~~\tilde B_{\alpha a}= {\dd \frac 1\omega} B_{\alpha a}\label{eq:81}.
\ee
These two types of contractions are not equivalent in the case of the Maxwell algebra, since this is not invariant under a full rescaling of the momenta. However, performing a rescaling of the charges by a factor $\omega^{-2}$ we recover the Maxwell algebra. We can easily show that there are no other equivalent solutions, except fo the ones found in  
Table 
\ref{table:2}, if we perform the scaling of eq. (\ref{eq:81}). In this case we define
\be
\tZ_{ab} =\omega^{t'} Z_{ab},~~~\tZ_{a\alpha}=\omega^{r'} Z_{a\alpha},~~~\tZ_{\alpha\beta}=\omega^{s'} Z_{\alpha\beta}.                           
\ee
Then, from the commutation rules of the scaled momenta we get:
\be
[\tP_a,\tP_b]= i  Z_{ab}=\frac i{\omega^{t'}}\tZ_{ab}\to t'\ge 0\label{eq:16},
\ee\be                                                                
[\tP_a,\tP_\alpha]= i\omega  Z_{a\alpha}=\frac i{\omega^{r'-1}}\tZ_{a\alpha}\to r'-1\ge 0\label{eq:17},
\ee
\be
[\tP_\alpha,\tP_\beta]= i\omega^2 Z_{\alpha\beta} =\frac i{\omega^{s'-2}}\tZ_{\alpha\beta}, \to s'-2\ge 0\label{eq:18}.
\ee
Translating all the exponents by 2:
\be
t' =t+2,~~~ r'=r +2,~~~s'=s+2,
\ee
we recover for $t,r,s$ the conditions given in eq. (\ref{eq:19a}). Considering that the scaling of the boosts is the same in the two cases we are considering, we obtain for the exponents $t',r',s'$ the same conditions we obtained for $t,r,s$ in eq. (\ref{eq:23}). Since these conditions depend only on the differences $t'-r'$ and $r'-s'$, we get the same result for the translated exponents.  This shows that also in this case we get the same 8 solutions given in Table \ref{table:2}, the only difference being the overall translation of the exponent by 2.

%
%
%
%
%

\end{document}